\documentclass[12pt]{article}
\pdfoutput=1
\usepackage{pstricks}
\usepackage{color}
\usepackage{amssymb,amsmath,bm,bbm}
\usepackage{epsf}
\usepackage{epsfig}
\usepackage{afterpage}
\usepackage{longtable}
\usepackage{cite}
\usepackage{latexsym,mathrsfs,dsfont}
\usepackage{graphics}
\usepackage{url}
\usepackage{paralist}
\usepackage{bbold}

\newcommand{\ord}{\mathcal{O}}

\newcommand{\tev}{\, {\rm TeV}}
\newcommand{\gev}{\, {\rm GeV}}
\newcommand{\mev}{\, {\rm MeV}}

\newcommand{\vcb}{|V_{cb}|}

\newcommand{\vub}{|V_{ub}|}

\newcommand{\bsi}{B_6^{(1/2)}}
\newcommand{\bei}{B_8^{(3/2)}}

\def\epe{\varepsilon'/\varepsilon}
\newcommand{\beq}{\begin{equation}}
\newcommand{\eeq}{\end{equation}}
\newcommand{\be}{\begin{equation}}
\newcommand{\ee}{\end{equation}}
\newcommand{\bi}{\begin{itemize}}
\newcommand{\ei}{\end{itemize}}
\newcommand{\ba}{\begin{array}}
\newcommand{\ea}{\end{array}}
\newcommand{\beqa}{\begin{eqnarray}}
\newcommand{\eeqa}{\end{eqnarray}}
\newcommand{\bea}{\begin{eqnarray}}
\newcommand{\eea}{\end{eqnarray}}
\newcommand{\beqn}{\begin{eqnarray}}
\newcommand{\eeqn}{\end{eqnarray}}

\newcommand{\D}{\Delta}

\definecolor{red}{cmyk}{0,1,1,0.4}

\usepackage{fancyhdr}
\advance \headheight by 3.0truept       % for 12pt mandatory...
\begin{document}

\begin{flushright}
    {FLAVOUR(267104)-ERC-118}\\
    {BARI-TH/16-705}
\end{flushright}

\medskip

\begin{center}
{\Large\bf
\boldmath{331 Models Facing the Tensions in $\Delta F=2$ Processes\\
with the Impact on  $\epe$, $B_s\to\mu^+\mu^-$ and $B\to K^*\mu^+\mu^-$}}
\\[0.8 cm]
{\large\bf Andrzej~J.~Buras$^{a,b}$ and Fulvia~De~Fazio$^{c}$ 
 \\[0.5 cm]}
{\small
$^a$TUM Institute for Advanced Study, Lichtenbergstr. 2a, D-85747 Garching, Germany\\
$^b$Physik Department, Technische Universit\"at M\"unchen,
James-Franck-Stra{\ss}e, \\D-85747 Garching, Germany\\
$^c$Istituto Nazionale di Fisica Nucleare, Sezione di Bari, Via Orabona 4,
I-70126 Bari, Italy}
\end{center}

\vskip0.41cm

%{\em Version of \today}

\begin{abstract}
\noindent
Motivated by the recently improved results from the Fermilab Lattice and MILC Collaborations  on the hadronic matrix elements entering  $\Delta M_{s,d}$ in
 $B_{s,d}^0-\bar B_{s,d}^0$ mixings and the resulting increased tensions 
between $\Delta M_{s,d}$  and $\varepsilon_K$ in the Standard Model (SM) and 
CMFV models, we demonstrate that these tensions 
can be removed in  331 models based on the gauge group $SU(3)_C\times SU(3)_L\times U(1)_X$ both for $M_{Z^\prime}$  in the LHC reach and well beyond it. 
But the implied new physics (NP) patterns in $\Delta F=1$ observables depend 
sensitively on the value of $\vcb$. Concentrating the analysis on three 331 models that have been selected by us previously on the basis of their performance in electroweak precision tests and $\epe$ we illustrate this for $\vcb=0.042$ and
 $\vcb=0.040$. We find that these new lattice data still allow for  positive shifts in $\epe$ up to $6\times 10^{-4}$ for
 $M_{Z^\prime}=3\tev$  and $\vub=0.0036$ for both values of $\vcb$ but for $M_{Z^\prime}=10\tev$ only for $\vcb=0.040$ such shifts can   be obtained.  {For 
$\vub=0.0042$ maximal shifts in $\epe$ increase to $\simeq 7\times 10^{-4}$.}
NP effects  in 
$B_s\to\mu^+\mu^-$ and in the Wilson coefficient $C_9$  are significantly larger  in all three models for the case of  $\vcb=0.040$.
In particular in two models  the rate for $B_s\to\mu^+\mu^-$ can be reduced by NP by $20\%$ for  $M_{Z^\prime}=3\tev$  resulting in values  in the ballpark of central values from CMS and LHCb. In the third
model a shift in $C_9$ up to $C_9^\text{NP}=-0.5$  is possible. For 
$\vcb=0.042$, NP effects in $B_s\to\mu^+\mu^-$ and in $C_9$ are by at least a factor of two smaller. For  $M_{Z^\prime}=10\tev$ NP effects in
$B_s\to\mu^+\mu^-$ and $C_9$, independently of $\vcb$, are 
at most at the level of a few percent. 
We also consider the simplest 331 model, analyzed recently in the literature,
in which $X=Y$, the usual hypercharge. We find that in 
this model NP effects in flavour observables are much smaller than in the 
three models with $X\not=Y$, in particular
NP contributions to the ratio $\epe$ are very strongly suppressed.   
Our analysis exhibits the important role of lattice QCD and of precise values 
of CKM parameters, in particular $\vcb$, for quark flavour phenomenology beyond 
the SM. It also demonstrates exceptional role of $\Delta F=2$ observables and 
of $\epe$ in testing high energy scales beyond the LHC.
\end{abstract}

\thispagestyle{empty}
\newpage
\setcounter{page}{1}

\tableofcontents

\section{Introduction}

The Standard Model (SM) describes globally the existing data on quark-flavour violating processes rather well \cite{Buras:2013ooa} but with the reduction 
of experimental errors and increased precision in non-perturbative and 
perturbative QCD and electroweak calculations a number of tensions
 at the level of $2-3\, \sigma$ seem to emerge in various seemingly unrelated 
observables. While some of these tensions could turn out to be the result of statistical 
fluctuations, underestimate of systematical and theoretical errors, it is not
excluded that eventually they all signal the presence of some kind of new physics (NP). Therefore, it is interesting to investigate what this NP could be.

In the present paper we will address some of these tensions in 
331 models based on the gauge group $SU(3)_C\times SU(3)_L\times U(1)_X$   \cite{Pisano:1991ee,Frampton:1992wt}. As these models have much smaller number of 
new parameters than supersymmetric models, Randall-Sundrum scenarios and 
Littlest Higgs models, it is not evident that they can remove all present 
tensions simultaneously. 

Our paper has been motivated by a recent analysis in \cite{Blanke:2016bhf}
 which demonstrates that 
the new lattice QCD results from Fermilab Lattice and MILC Collaborations 
  \cite{Bazavov:2016nty}   on $B^0_{s,d}-\bar B^0_{s,d}$ hadronic matrix elements imply 
a significant tension between  
 $\varepsilon_K$ and $\Delta M_{s,d}$ within the SM. The authors of 
 \cite{Bazavov:2016nty} find also inconsistences
between $\Delta M_{s,d}$ and tree-level determination of $\vcb$. But the 
simultaneous consideration of  $\varepsilon_K$ and $\Delta M_{s,d}$ 
in \cite{Blanke:2016bhf} also demonstrates that the tension between these two 
quantities cannot be removed for any value of $\vcb$. Moreover, the situation worsens for other 
 models with constrained MFV (CMFV), indicating the presence of new flavour- and CP-violating interactions beyond CMFV framework at work.

The question then arises how  
331 models face this tension and what are the implications 
of new lattice results on other observables for which some departures from 
SM predictions have been identified. 
In particular, taking the results in \cite{Blanke:2016bhf,Bazavov:2016nty} into  account we want to concentrate our analysis on 
\be
\epe, \qquad  B_s\to\mu^+\mu^-, \qquad B\to K^*\mu^+\mu^-\,.
\ee

In this context the following facts should be recalled.
\begin{itemize}
\item
Recent analyses in  \cite{Blum:2015ywa,Bai:2015nea,Buras:2015yba,Buras:2015xba} 
find the ratio $\epe$ in the SM to be significantly  below  
the experimental world average from NA48 \cite{Batley:2002gn} and KTeV
\cite{AlaviHarati:2002ye,Abouzaid:2010ny} collaborations.
The recent analysis in the large $N$ approach in \cite{Buras:2016fys}
 indicates  that final state interactions will not modify this picture at least 
on the qualitative level. The analysis in \cite{Buras:2015yca} shows 
that CMFV models cannot cure this problem. In any case models providing
 an enhancement of $\epe$ should be favoured from present perspective.
\item
The branching ratio for $B_s\to\mu^+\mu^-$ measured by CMS and LHCb  \cite{CMS:2014xfa} has been always 
 visibly below rather precise prediction of the  SM \cite{Bobeth:2013uxa}.  The most recent result from ATLAS\footnote{$\overline{{B}}(B_s\to\mu^+\mu^-)=(0.9^{+1.1}_{-0.9})\times 10^{-9}$ \cite{Aaboud:2016ire}.},
 while not accurate, appears to confirm this picture and models suppressing the rate for this decay relative to its SM prediction appear to be  favoured. 
\item
LHCb data on  $B_d\to K(K^*)\mu^+\mu^-$ indicate some departures from SM 
expectation although this issue is controversial. See \cite{ Altmannshofer:2014rta,Descotes-Genon:2015uva} and references to the rich literature therein. Assuming again that statistical fluctuations 
or underestimated errors are not responsible for these effects, significant 
NP contributions to the Wilson coefficient $C_9$ or $C_9$ and $C_{10}$ 
are required.
\end{itemize}

These three items have been already addressed by us within 331 models 
in the past \cite{Buras:2012dp,Buras:2013dea,Buras:2014yna,Buras:2015kwd}.
In particular in \cite{Buras:2015kwd} the issue of $\epe$ anomaly has been 
addressed, while in \cite{Buras:2013dea,Buras:2015kwd} the last two items above have been considered. The main result of \cite{Buras:2015kwd} is that among
 24 331 models only three (M8, M9 and M16 in the terminology of \cite{Buras:2014yna}) have a chance to survive if an improved 
fit to electroweak precision observables relative to the SM is required and the $\epe$ anomaly 
will be confirmed in the future. Two of them (M8 and M9)  allowed simultaneously a suppression of the 
rate for   $B_{s}\to \mu^+\mu^-$ by $20\%$ thereby bringing the  theory closer to the data without any significant impact 
on the Wilson coefficient $C_9$.  The third model (M16) provided, simultaneously  to the enhancement of $\epe$, a 
shift  up to $\Delta C_9=-0.6$, softening the anomalies in $B\to K^*\mu^+\mu^-$, without any significant impact on  $B_{s}\to \mu^+\mu^-$. While, before the ATLAS data on $B_s\to \mu^+\mu^-$, M16 seemed to be slightly favoured over M8 and M9, this data and the fact that the theoretical uncertainties in  $B_{s}\to \mu^+\mu^-$ are significantly smaller than in 
 $B_d\to K^*\mu^+\mu^-$  make us believe that at the end models M8 and M9 have a bigger chance to survive.

However, the constraints from $\Delta F=2$ transitions used in \cite{Buras:2015kwd},
prior to the lattice QCD result in \cite{Bazavov:2016nty}, were significantly weaker
 and it is of interest to investigate what is the impact of these new 
lattice results on our previous analyses and whether the increased tension between 
 $\varepsilon_K$ and $\Delta M_{s,d}$ within the SM pointed out in 
 \cite{Blanke:2016bhf} can be removed in these three models. In fact one 
should recall that the mixing and CP-violation in $B^0_{s,d}-\bar B^0_{s,d}$ 
systems play very important roles in the determination of new parameters 
in  331 models \cite{Buras:2012dp} and it is not surprizing that our previous results will be indeed
modified in a visible manner.

 In this context let us remark that within the SM, 
dependently on whether $\Delta M_{s}$ or $\varepsilon_K$ has been used as a constraint, rather different values for $\vcb$ have been required to fit the data within the SM \cite{Blanke:2016bhf}:
\be\label{VCB}
\vcb=(39.7\pm1.3)\times 10^{-3}\quad (\Delta M_s),\qquad \vcb=(43.3\pm1.1)\times 10^{-3} \quad (\varepsilon_K).
\ee
This in turn resulted in rather different predictions for rare $K$ and $B_{s,d}$  decays as seen in Table~4 of  \cite{Blanke:2016bhf}. 

In our most recent analysis in \cite{Buras:2015kwd} we have performed numerical
analysis for $\vcb$ in the ballpark of the higher value in (\ref{VCB}), that 
is $0.042$. In the present paper we will also use this value 
in order to see the impact of new lattice data on our previous results, but in addition we will 
perform the analysis with  $0.040$ which is in the ballpark
of its lower value in (\ref{VCB}).  This will tell us whether 331 models can 
cope with the tensions in question for both values of $\vcb$ and whether 
the implications for NP effects are modified through this change of $\vcb$.

The second motivation for a new analysis of 331 models is the following 
one. In our analyses and also in \cite{Diaz:2004fs,CarcamoHernandez:2005ka,Promberger:2007py} the $U(1)_X$ factor in the  gauge group $SU(3)_C\times SU(3)_L\times U(1)_X$   differed from   the hypercharge gauge group $U(1)_Y$.
As various 331 models are characterized by 
two parameters $\beta$ and $\tan\bar\beta$ defined through 
\be\label{QTX}
Q=T_3+\frac{Y}{2}= T_3+\beta T_8+X, \qquad \tan\bar\beta=\frac{v_\rho}{v_\eta}\,
\ee
these analyses dealt with $\beta\not=0$.
Here $T_{3,8}$ and $X$ are the  diagonal generators of  $SU(3)_L$ and 
$U(1)_X$, respectively. $Y$ represents $U(1)_Y$ and $v_i$ are  
the vacuum expectation values of scalar triplets responsible for the generation of down- and up-quark masses in these models.

Recently a special variant of 331 models with $\beta=0$ or equivalently 
$U(1)_X=U(1)_Y$ has been considered in \cite{Hue:2015mna}. Moreover, these authors 
set $\tan\bar\beta=1$ as this choice with $\beta=0$ simplifies the model by 
eliminating  $Z-Z^\prime$ mixing studied by us in detail in \cite{Buras:2014yna} for $\beta\not=0$. As this is the simplest among the 331 models, the 
question  arises whether it is consistent with the flavour data in 
the setup in \cite{Hue:2015mna} and what are the
 implications for quark flavour observables for arbitrary  $\tan\bar\beta$ 
when $Z-Z^\prime$ mixing enters the game. In particular  the comparison with our studies for $\beta\not=0$ in \cite{Buras:2012dp,Buras:2013dea,Buras:2014yna,Buras:2015kwd} and in the present paper is of interest. As the authors of  \cite{Hue:2015mna} did not address this question, to our knowledge this is the first quark flavour study of this simplified 331 model. 

We will see that in the absence of $Z-Z^\prime$ mixing 
 the choice $\beta=0$ provides a unique 331 model in 
which the phenomenologically successful relation
\be\label{R1}
C_9^\text{NP}=-C_{10}^\text{NP}
\ee
is satisfied.  Here $C_9^\text{NP}$ and $C_{10}^\text{NP}$ stand for the shifts in the 
Wilson coefficients relevant in particular for $B\to K^*\mu^+\mu^-$ and 
$B_s\to\mu^+\mu^-$, respectively. This is good news. The bad news 
is that setting $\beta=0$ modifies the values of all couplings relative to the ones in M8, M9 and M16 models. We find then that NP contributions to $\epe$ in 
this simple model are at most $1 \times 10^{-4}$ for  $M_{Z^\prime}=3\tev$ and
decrease with increasing  $M_{Z^\prime}$.
 The effects in $C_9^\text{NP}$ and $C_{10}^\text{NP}$ are at most  at the level of a few percent even if $Z-Z^\prime$ mixing is taken into 
account. Thus the model fails in solving three anomalies listed above.
But as we will see it is able to remove
 the  tensions between $\Delta M_{s,d}$ and $\varepsilon_K$.

Our paper is organized as follows.
In Section~\ref{sec:2} we address the  tensions between 
$\Delta M_{s,d}$ and $\varepsilon_K$ in  M8, M9 and M16 models and 
we update  our analysis of $\epe$, $B_s\to\mu^+\mu^-$ and $C_9$ in  \cite{Buras:2015kwd}  taking new $\Delta F=2$ constraints from \cite{Bazavov:2016nty}
into account and performing the analysis at two values of $\vcb$ as discussed 
above.
In Section~\ref{sec:3} we specify the existing formulae in 331 models to 
the case $\beta=0$ but for arbitrary $\tan\bar\beta$ and we derive the results 
mentioned above. We conclude in Section~\ref{sec:4}.

\section{M8, M9 and M16 Facing Anomalies}\label{sec:2}
\subsection{Preliminaries}
Let us recall that in these three models new flavour-violating effects are 
governed by tree-level $Z^\prime$ exchanges with a subdominant role played 
by tree-level $Z$ exchanges generated through $Z-Z^\prime$ mixing. All the 
formulae for flavour observables in these models can be found in 
 \cite{Buras:2012dp,Buras:2013dea,Buras:2014yna,Buras:2015kwd} and will not 
be repeated here. In particular the collection of formulae for $Z^\prime$ couplings to quarks and leptons for arbitrary $\beta$
are given in (17) and (18) of \cite{Buras:2013dea}.

New sources of flavour and CP violation in 331 models are parametrized by
new mixing parameters and phases
\be\label{PAR}
\tilde s_{13},\qquad\tilde s_{23},\qquad  \delta_1,\qquad \delta_2
\ee
with $\tilde s_{13}$ and $\tilde s_{23}$ positive definite  and smaller than  unity and 
$0\le \delta_{1,2}\le 2\pi$. They can be constrained by flavour observables as demonstrated in detail in \cite{Buras:2012dp}. 
The non-diagonal $Z^\prime$ couplings  relevant for $K$, $B_d$ and $B_s$ meson 
systems can 
be then parametrized respectively within an excellent approximation through
\be\label{vij}
 v_{32}^*v_{31}=\tilde s_{13}\tilde s_{23}e^{i(\delta_2-\delta_1)}, \qquad
 v_{33}^*v_{31}=-\tilde s_{13}e^{-i\delta_1}, \qquad
 v_{33}^*v_{32}=-\tilde s_{23}e^{-i\delta_2} \,.
\ee
$\tilde s_{13}$ and $\delta_1$ can be determined from $\Delta M_d$ and CP-asymmetry $S_{\psi K_S}$ while $\tilde s_{23}$ and $\delta_2$ from $\Delta M_s$ and CP-asymmetry $S_{\psi \phi}$. Then the parameters in the $K$ system are fixed. This correlation tells us that the removal of tensions between $\varepsilon_K$ and $\Delta M_{s,d}$ is not necessarily  automatic in 331 models and constitutes an important test of these models.

The remaining two parameters, except for $M_{Z^\prime}$ mass, are as seen in 
(\ref{QTX}), $\beta$ and $\tan\bar\beta$. Moreover, the fermion representations 
of SM quarks under the $SU(3)_L$ group matter. The three models in question 
are then characterized by
\begin{align}
\beta&=\frac{2}{\sqrt{3}},\qquad  \tan\bar\beta= 5\, \qquad (F_1), \qquad ({\rm M8}),\\
\beta&=-\frac{2}{\sqrt{3}},\quad  \,\tan\bar\beta= 1\, \qquad (F_2), \qquad ({\rm M9}),\\
\beta&=\frac{2}{\sqrt{3}},\qquad  \tan\bar\beta= 5\, \qquad (F_2), \qquad ({\rm M16})\,
\end{align}
with $F_1$ and $F_2$ standing for two fermion representations. In $F_1$ 
the first two generations of quarks belong to triplets of $SU(3)_L$,
while the third generation of quarks to an antitriplet. In $F_2$ it 
is opposite. 
With the values of  $\beta$ and $\tan\bar\beta$ being fixed flavour phenomenology depends only on the parameters in (\ref{PAR}) and  $M_{Z^\prime}$.

\begin{figure}[!tb]
\centering
\includegraphics[width = 0.47\textwidth]{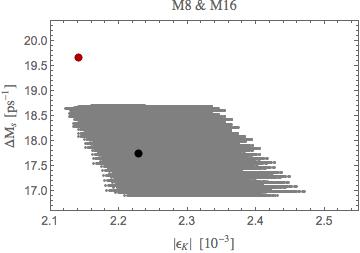}
\includegraphics[width = 0.47\textwidth]{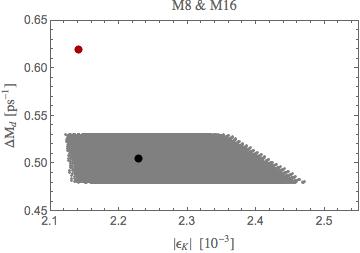}
 \caption{\it  $\Delta M_{s,d}$ vs.  $\varepsilon_K$ in M8 and M16. 
 Red dots represent central SM values, black dots the central experimental values. $M_{Z^\prime}=3\tev$ and $\vcb=0.042$.}\label{MBAJB}
\end{figure}

\subsection{Numerical Analysis}
The difficulty in doing the numerical analysis are  tensions between 
inclusive and exclusive determinations of the CKM elements $\vcb$ and $\vub$.
The exclusive  determinations have been summarized 
in \cite{DeTar:2015orc} and are given as follows
\be\label{excl}
\vcb_\text{excl}=(39.78\pm0.42)\cdot 10^{-3},\qquad \vub_\text{excl}=(3.59\pm0.09)\cdot 10^{-3}.
\ee
They are based on \cite{Lattice:2015rga,Lattice:2015tia,Glattauer:2015teq,Bazavov:2016nty,Aaij:2015bfa}.
The inclusive ones are summarized well in 
\cite{Alberti:2014yda,Gambino:2016fdy}
\be\label{incl}
\vcb_\text{incl}=(42.21\pm0.78)\cdot 10^{-3},\qquad \vub_\text{incl}=(4.40\pm0.25)\cdot 10^{-3}.
\ee
We note that after the recent Belle data on $B\to D\ell\nu_l$ \cite{Glattauer:2015teq}, the exclusive and inclusive values of $\vcb$ are closer to each other 
than in the past. On the other hand in the case of $\vub$ there is a very significant difference. 

Furthermore, after recent precise determinations of hadronic matrix elements 
 entering  $\Delta M_{s,d}$ in  $B_{s,d}^0-\bar B_{s,d}^0$ mixing by 
 Fermilab Lattice and MILC Collaborations \cite{Bazavov:2016nty} there 
are significant tensions between tree-level determinations of $\vcb$ and $\vub$
and $\Delta M_{s,d}$ within the SM  \cite{Bazavov:2016nty}  and also 
the tensions between $\varepsilon_K$ and $\Delta M_{s,d}$ \cite{Blanke:2016bhf}
in this model. Moreover, as found in the latter paper, the value of the angle $\gamma$ in the unitarity 
triangle extracted from the ratio $\Delta M_d/\Delta M_s$ and the CP-asymmetry
 $S_{\psi K_S}$ is with $\gamma=(63.0\pm 2.1)^\circ$ visibly smaller than it 
tree-level determination
\cite{Trabelsi:2014}
\begin{equation}\label{gamma}
    \gamma = (73.2^{+6.3}_{-7.0})^\circ.
\end{equation}

In the present paper, as in \cite{Buras:2015kwd}, we will
set  first the CKM parameters to
\be\label{CKMfix}
\vub=3.6\times 10^{-3}, \qquad \vcb=42.0 \times 10^{-3}, \qquad \gamma=70^\circ.
\ee
This choice is in the ballpark  of exclusive determination of $\vub$ in (\ref{excl}) and 
the inclusive one for $\vcb$ in (\ref{incl}).  Moreover, it is in the ballpark of tree-level determination 
of $\gamma$. In view of new parameters in 331 models the value of $\gamma$ does not follow
from the ratio $\Delta M_s/\Delta M_d$ and $S_{\psi K_S}$ like in CMFV models 
and it is better to take $\gamma$ from tree-level determinations as it is
to first approximation not polluted by NP.  Having the same CKM input
as in our previous analysis will allow us to see the impact of new lattice
data on phenomenology.

The choice in (\ref{CKMfix}) is also motivated by the fact that NP contributions to 
$\varepsilon_K$ in 331 models are rather small for $M_{Z^\prime}$ of a few $\tev$ and SM 
should perform well in this case. Indeed for this choice of CKM parameters 
we find
\be
|\varepsilon_K|_{\rm SM}=2.14\times 10^{-3}, \qquad (\Delta M_K)_{\rm SM}=0.467\cdot 10^{-2} \,\text{ps}^{-1}
\ee
and  $|\varepsilon_K|$ in 
the SM only $4\%$ below the data. Due to the presence of long distance effects
in $\Delta M_K$ also this value is compatible with the data.

 While the  CKM parameters do not enter 
 the shift in $\epe$ and $\varepsilon_K$, their choice  matters in the predictions for NP contributions to $\Delta F=2$ observables in $B_{d,s}^0-\bar B_{d,s}^0$ 
systems and the rare $B_{s,d}$ decays. This is not only because 
of their interferences with  SM  contributions. The departure of SM 
predictions for $\varepsilon_K$ and $\Delta M_{s,d}$ from the data depends on 
the CKM parameters, in particular on the value of $\vcb$, and this has an
 impact on the allowed ranges of new parameters extracted from 
$\Delta F=2$ observables and consequently on final values of $\epe$, 
$\overline{\mathcal{B}}(B_s\to\mu^+\mu^-)$ and the shift in $C_9$. We will
illustrate this below by choosing also $\vcb=0.040$ 
  which corresponds to its exclusive determination in 
(\ref{excl}). See (\ref{CKMfix1}).

\begin{figure}[!tb]
 \centering
\includegraphics[width = 0.47\textwidth]{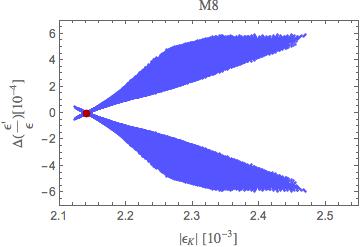}
\includegraphics[width = 0.47\textwidth]{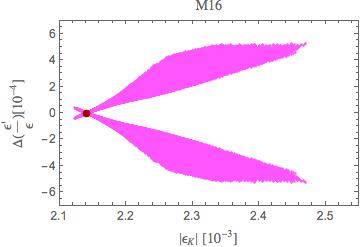}
\caption{ \it $\Delta(\epe)$ versus $\varepsilon_K$ for M8 and M16.
 Red dots represent central SM values.  $M_{Z^\prime}=3\tev$ and $\vcb=0.042$.}\label{M8M16}~\\[-2mm]\hrule
\end{figure}

\begin{figure}[!tb]
 \centering
\includegraphics[width = 0.47\textwidth]{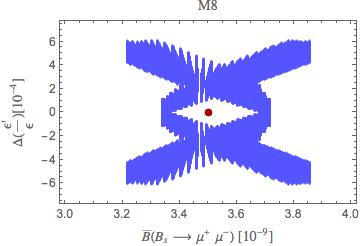}
\includegraphics[width = 0.47\textwidth]{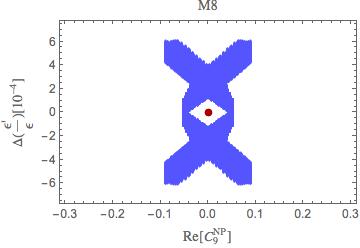}
\includegraphics[width = 0.47\textwidth]{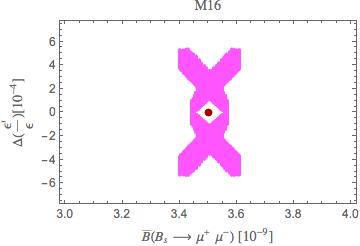}
\includegraphics[width = 0.47\textwidth]{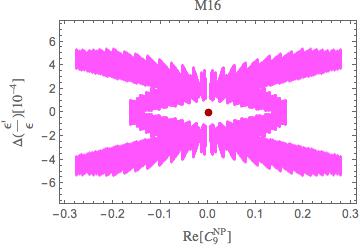}
\caption{ \it Correlations of $\Delta(\epe)$ with $B_s\to \mu^+\mu^-$ 
(left panels) and with $ C_9^{\rm NP}$ (right panels) for M8 and M16.
Red dots represent central SM values. 
$M_{Z^\prime}=3\tev$ and $\vcb=0.042$.}
\label{CORR816}~\\[-2mm]\hrule
\end{figure}

Next, as in \cite{Buras:2012dp,Buras:2015kwd}, we  perform a simplified analysis of
$\Delta M_{d,s}$, $S_{\psi K_S}$
and $S_{\psi\phi}$
in order to identify oases in the space of four parameters (\ref{PAR})
for which these four observables are consistent with experiment.
To this end we use the formulae for $\Delta F=2$ observables in 
 \cite{Buras:2012dp,Buras:2014yna} and  set input parameters listed in 
Table~3 of our recent analysis in  \cite{Buras:2015kwd} at their central values.  The only modifications in this input  are the recently calculated  parameters  \cite{Bazavov:2016nty} \footnote{These results are more accurate than ETM results \cite{Carrasco:2013zta}, but compatible with them. We look forward to new FLAG averages on these 
quantities.} 
\be\label{Kronfeld}
 F_{B_s}\sqrt{\hat B_{B_s}}=(274.6\pm8.8)\mev,\qquad  F_{B_d} \sqrt{\hat B_{B_d}}=
(227.7\pm 9.8)\mev \,,
\ee
that should be compared with $F_{B_s}\sqrt{\hat B_{B_s}}=(266.0\pm18.0)\mev$ and 
$F_{B_d}\sqrt{\hat B_{B_d}}=(216.0\pm 15.0)\mev$ used by us in \cite{Buras:2015kwd} . This change implies the modifications in the SM values of $\Delta M_{s,d}$ 
that now are significantly higher than the data: 
\be
 (\Delta M_s)_{\rm SM}=19.66/{\rm ps},\qquad (\Delta M_d)_{\rm SM}=0.620/{\rm ps},
\qquad S^{\rm SM}_{\psi\phi}=0.037, \qquad S^{\rm SM}_{\psi K_S}=0.688\,
\ee
with CP asymmetries unchanged and compatible with the data. Thus the 331 
models are requested to bring the values of $\Delta M_{s,d}$ down to their
experimental values \cite{Amhis:2014hma}
\be
 (\Delta M_s)_{\rm exp}=17.757(21)/{\rm ps},\qquad (\Delta M_d)_{\rm exp}=0.5055(20)/{\rm ps},
\ee
while being consistent with the data for $\varepsilon_K$,  $S_{\psi K_S}$ and $S_{\psi\phi}$.

As we keep the input parameters at their central values, in order to take partially hadronic
and experimental uncertainties into account we require the 331 models
to reproduce the data for $\Delta M_{s,d}$ within $\pm 5\%$ 
and the
data on $S_{\psi K_S}$ and $S_{\psi\phi}$ within experimental
$2\sigma$ ranges.

Specifically, our search is governed by the following allowed ranges:
\be
16.9/{\rm ps}\le \Delta M_s\le 18.7/{\rm ps},
\qquad  -0.055\le S_{\psi\phi}\le 0.085, \label{oases23}
\ee
\be
0.48/{\rm ps}\le \Delta M_d\le 0.53/{\rm ps},\qquad  0.657\le S_{\psi K_S}\le 0.725\, . \label{oases13}
\ee
The ranges for $\Delta M_{s,d}$ are smaller than used in  \cite{Buras:2015kwd}
because of the reduced errors in (\ref{Kronfeld}).

{We also impose the constraint on the ratio $\Delta M_s/\Delta M_d$ using 
 \cite{Bazavov:2016nty} 
\be\label{xi}
\xi=\frac{F_{B_s}\sqrt{\hat B_{B_s}}}{F_{B_d}\sqrt{\hat B_{B_d}}}=1.206\pm0.019\,.
\ee
In the spirit of our simplified analysis we will keep this ratio 
at its central value in (\ref{xi}) but in order to take into account the uncertainty in $\xi$ we will 
require that $\Delta M_s/\Delta M_d$ agrees with the data 
within $\pm 5\%$. Specifically we will require that
\be\label{RCMFV}
33.3\le \left(\frac{\Delta M_s}{\Delta M_d}\right)\le 36.8\,
\ee
is satisfied.}

In the case of $\varepsilon_K$  and $\Delta M_K$ we will just proceed as 
in  \cite{Buras:2015kwd} imposing the ranges
\be\label{CONE}
1.60 \times 10^{-3}< |\varepsilon_K|< 2.50\times 10^{-3}\,, \qquad
-0.30\le \frac{(\Delta M_K)^{Z^\prime}}{(\Delta M_K)_\text{exp}}\le 0.30\,.
\ee

Having determined the ranges for the parameters (\ref{PAR}) we can calculate 
all the remaining flavour observables of interest.

\begin{figure}[!tb]
\centering
\includegraphics[width = 0.47\textwidth]{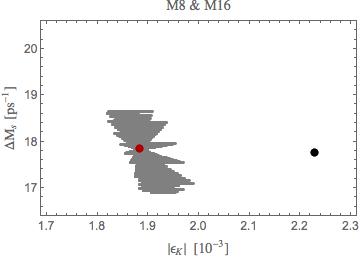}
\includegraphics[width = 0.47\textwidth]{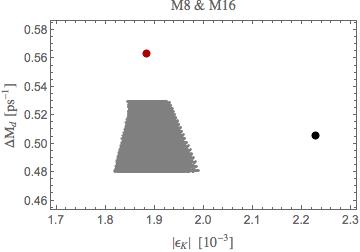}
 \caption{\it  $\Delta M_{s,d}$ vs.  $\varepsilon_K$ in M8 and M16.
 Red dots represent central SM values, black dots the central experimental values. $M_{Z^\prime}=3\tev$, $\vcb=0.040$ and $\vub=0.0036$.}\label{MBAJBv4}
\end{figure}

\begin{figure}[!tb]
 \centering
\includegraphics[width = 0.47\textwidth]{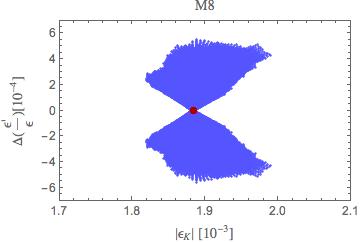}
\includegraphics[width = 0.47\textwidth]{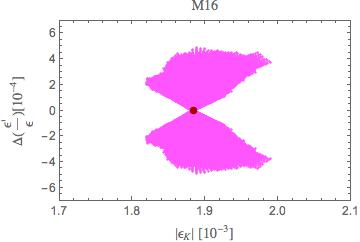}
\caption{ \it $\Delta(\epe)$ versus $\varepsilon_K$ for M8 and M16. Red dots represent central SM values.  $M_{Z^\prime}=3\tev$, $\vcb=0.040$ and $\vub=0.0036$.}\label{M8M16v4}~\\[-2mm]\hrule
\end{figure}

\begin{figure}[!tb]
 \centering
\includegraphics[width = 0.47\textwidth]{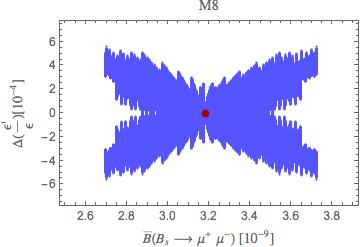}
\includegraphics[width = 0.47\textwidth]{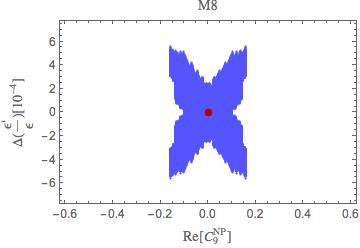}
\includegraphics[width = 0.47\textwidth]{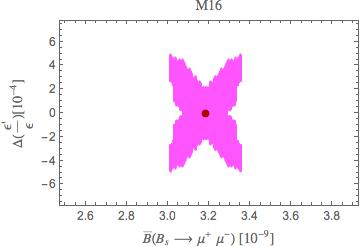}
\includegraphics[width = 0.47\textwidth]{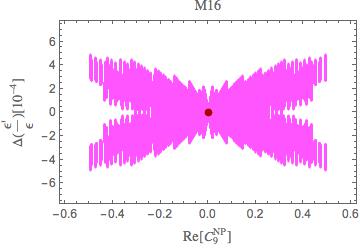}
\caption{ \it Correlations of $\Delta(\epe)$ with $B_s\to \mu^+\mu^-$ 
(left panels) and with $ C_9^{\rm NP}$ (right panels) for M8 and M16. 
 Red dots represent central SM values.
 $M_{Z^\prime}=3\tev$, $\vcb=0.040$ and $\vub=0.0036$.}
\label{CORR816vcb4}~\\[-2mm]\hrule
\end{figure}

In Fig.~\ref{MBAJB} we show $\Delta M_{s,d}$ vs. $\varepsilon_K$ in M8 and M16. Red dots represent central SM values and black dots the central experimental 
values.  The experimental errors are negligible and the parametric and theoretical errors are represented by the allowed departure from them as explained 
above. These results do not depend on the fermion representation up to tiny 
effects from $Z-Z^\prime$ mixing and consequently are practically the same 
for M8 and M16.  As far as  M9 is concerned all results presented in our paper
 are very similar to the ones in M8 and will not be shown. We observe that 
the tensions between   $\Delta M_{s,d}$ vs.  $\varepsilon_K$ present in the 
SM  can be easily removed in 331 models for $M_{Z^\prime}=3\tev$.

In Fig.~\ref{M8M16} we show $\Delta(\epe)$ versus $\varepsilon_K$ for M8 and M16
 at $M_{Z^\prime}=3\tev$. Taking the uncertainties due to charm contribution and 
 CKM parameters in $\varepsilon_K$ into account the maximal shifts in $\epe$ 
in both models amount to 
$6\times 10^{-4}$, very similar to what we found in \cite{Buras:2015kwd}.
But NP effects in $B_s\to\mu^+\mu^-$ and $C_9$ are  smaller relative 
to the ones found in the latter paper by a factor of two.
This is seen in  Fig.~\ref{CORR816}, where
we show correlations of $\Delta(\epe)$ with $B_s\to \mu^+\mu^-$ 
(left panels) and with $ C_9^{\rm NP}$ (right panels) for M8 and M16 and
$M_{Z^\prime}=3\tev$.  

In M8 the rate for $B_s\to\mu^+\mu^-$ can be suppressed 
by $10\%$ bringing the theory closer to the data in (\ref{LHCb2})
\cite{CMS:2014xfa}. Moreover, this happens
for the largest shift in $\epe$. But the shift in $C_9$ is very small. In 
M16 the pattern is opposite with only a very small NP effects in 
$B_s\to\mu^+\mu^-$ and a shift of $-0.3$ in $C_9$ which brings the theory 
closer to the data.
\boldmath
\subsection{$\vcb$ and $\vub$ Dependence}
\unboldmath
It is well known that $\varepsilon_K$ and $\Delta M_{s,d}$  in the SM are sensitive functions of $\vcb$. Moreover,  $\varepsilon_K$ and $S_{\psi K_S}$ depend sensitively on $\vub$.
 Setting 
$\vcb$ and $\vub$ to the values in (\ref{CKMfix}) we have necessarily constrained  the allowed range of NP parameters that are consistent with the data on 
$\Delta F=2$ observables. Changing $\vcb$ and $\vub$ will necessarily modify 
this range and will modify NP contributions to flavour observables even if they  do not depend directly on $\vcb$ and $\vub$.
A sophisticated analysis 
which would include the uncertainties in both CKM elements from tree-level decays would wash out NP effects and 
would not teach us much about the impact of  $\vcb$ and $\vub$  on our results.

 Therefore, 
we prefer to show how our results presented above are modified for 
a different value of $\vcb$ that we choose to be lower so that instead of 
 (\ref{CKMfix}) we now use
\be\label{CKMfix1}
\vub=3.6\times 10^{-3}, \qquad \vcb=40.0 \times 10^{-3}, \qquad \gamma=70^\circ.
\ee 
We keep $\vub$ and $\gamma$ unchanged as this will allow us to see the 
role of $\vcb$ better. The dependence on $\gamma$ in the observables in question is weak. The dependence of $\Delta M_{s,d}$ on $\vub$ is totally negligible. The inclusive value of $\vub$ would compensate the decrease of $\vcb$
in $\varepsilon_K$  but would simultaneously have an impact on $S_{\psi K_S}$ 
shifting it in the ballpark of $0.80$ within the SM. While in the SM this 
problem cannot be cured because of the absence of new CP-violating phases, in 
331 models the presence of the phase $\delta_1$ in (\ref{vij}) allows to
satisfy the constraint on $S_{\psi K_S}$ in (\ref{oases13}).  We will demonstrate it below.

This assures us that the tension between $\varepsilon_K$ and $S_{\psi K_S}$
for exclusive value of $\vcb$ present in the SM can be avoided within the 
331 models. But as we would like to investigate the impact of the change in $\vcb$ on our results, we keep first $\vub$ at its exclusive value. Moreover
 there is some
kind of consensus in the community that in the case of $\vub$ one can trust 
more exclusive determinations of this parameter than the inclusive ones. This 
is based on the fact that the exclusive determinations use formfactors from 
lattice QCD, which on the one hand are already rather precise and on the other hand do not require
the assumptions like hadron duality necessary for the inclusive determination 
of $\vub$.

\begin{figure}[!tb]
\centering
\includegraphics[width = 0.47\textwidth]{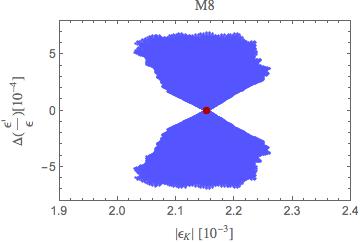}
 \caption{\it  $\Delta(\epe)$ versus $\varepsilon_K$ for M8.
 Red dot represents central SM value. 
$M_{Z^\prime}=3\tev$, $\vcb=0.040$ and $\vub=0.0042$.}\label{epeVubH3TeV}
\end{figure}

The 
results of this exercise are shown in Figs.~\ref{MBAJBv4}-\ref{CORR816vcb4}.
Comparing Fig.~\ref{MBAJBv4} with Fig.~\ref{MBAJB} it is evident that 331 models
perform better for our nominal choice of CKM parameters in (\ref{CKMfix}) than 
for a lower value of $\vcb$. This is seen in particular in the case of $\varepsilon_K$ for which the maximal values of $\varepsilon_K$ are by $10\%$ below the
data.  But, as discussed above, increasing $\vub$ towards its inclusive value and taking the uncertainties in the QCD corrections to the charm contribution in $\varepsilon_K$ into account one can bring the theory much closer to data  without
 violating the constraint on $S_{\psi K_S}$ in (\ref{oases13}).  We demonstrate this in Fig.~\ref{epeVubH3TeV} where we use $\vub=0.0042$. Indeed $\varepsilon_K$ is now in a perfect agreement with the data. The slight increase of the 
maximal value of $\Delta(\epe)$ in this case will be analyzed in more details below.

Therefore we can claim that 331 
models also in this case remove the tensions in question, which is not possible
 within the SM. Interestingly, as we will see below for significantly 
higher values of  $M_{Z^\prime}$ the removal of tensions for this value of 
$\vcb$ will be much easier. We refer to Section~4 in \cite{Buras:2015kwd} for the explanation of this behaviour.

Next Figs.~\ref{M8M16v4} and \ref{CORR816vcb4} should be compared with 
 Figs.~\ref{M8M16} and \ref{CORR816}, respectively. We observe:
\begin{itemize}
\item
The correlation between $\epe$ and $\varepsilon_K$ in  Fig.~\ref{M8M16v4}
has a very different shape than in  Fig.~\ref{M8M16} but a shift of $\epe$ of $(5-6)\times 10^{-4}$ is 
possible. In fact this plot is similar to a corresponding plot in  \cite{Buras:2015kwd} obtained with CKM parameters in (\ref{CKMfix}) but older hadronic matrix
elements. This similarity is easy to understand. The increase of non-perturbative parameters in (\ref{Kronfeld}) has been roughly compensated by the decrease
of $\vcb$.
\item
The size of NP effects in $B_s\to\mu^+\mu^-$ and $C_9$ is now larger than 
for our nominal value of $\vcb$ and similar to the ones found in \cite{Buras:2015kwd}: suppression of the rate for $B_s\to\mu^+\mu^-$ by $20\%$ in the case of
M8 and a shift of $C_9$ by $-0.5$ in M16 are possible. But what is interesting is that the decreased value of $\vcb$ lowers also the SM result for the  $B_s\to\mu^+\mu^-$ rate so that with the NP shift central values from CMS and LHCb 
\cite{CMS:2014xfa}
\be\label{LHCb2}
\overline{{B}}(B_{s}\to\mu^+\mu^-) = (2.8^{+0.7}_{-0.6}) \cdot 10^{-9}, 
\ee
can be reached. 
\end{itemize}
This value should be compared with central SM values 
\be
\overline{{B}}(B_{s}\to\mu^+\mu^-)_\text{SM} = 3.5 \cdot 10^{-9},\qquad
\overline{{B}}(B_{s}\to\mu^+\mu^-)_\text{SM} = 3.2 \cdot 10^{-9}
\ee
for $\vcb=0.042$ and $\vcb=0.040$, respectively.
Thus within 331 models, on the whole, the results for $\Delta F=1$ for $\vcb=0.040$ appear more interesting than for $\vcb=0.042$. As we will see below 
this is in particular the case for larger values of  $M_{Z^\prime}$. 

\begin{figure}[!tb]
\centering
\includegraphics[width = 0.47\textwidth]{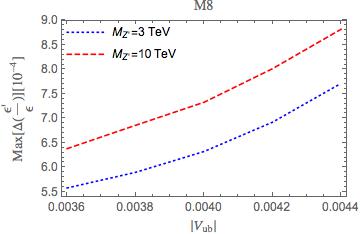}
 \caption{\it  Maximal values of $\Delta(\epe)$ for $\vcb=0.040$ as 
function of $\vub$ for $M_{Z^\prime}=3\tev$ and $M_{Z^\prime}=10\tev$.}\label{MBAJB10TeV}
\end{figure}

 Finally, we show in Fig.~\ref{MBAJB10TeV} the maximal value of $\Delta(\epe)$ for $\vcb=0.040$ as a
function of $\vub$. We observe that this value rises approximately linearly with increasing 
$\vub$ and for  $M_{Z^\prime}=3\tev$ and $\vub=0.0044$ that is consistent with 
the inclusive determinations could reach values as high as $\simeq 7.7\times 10^{-4}$. 
 This possibility should be kept in mind even if such high values of $\vub$ 
seem rather unlikely as stated above. For $\vcb=0.042$ the effects of changing 
$\vub$ turn out to be smaller.

\begin{figure}[!tb]
\centering
\includegraphics[width = 0.47\textwidth]{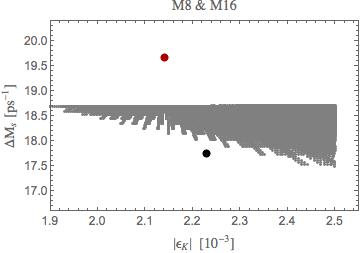}
\includegraphics[width = 0.47\textwidth]{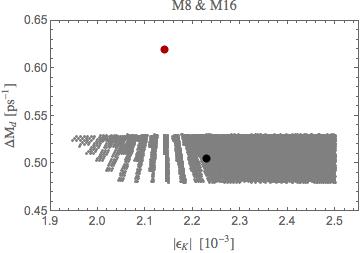}
 \caption{\it  $\Delta M_{s,d}$ vs.  $\varepsilon_K$ in M8 and M16. 
 Red dots represent central SM values and black dots the central experimental values. $M_{Z^\prime}=10\tev$, $\vcb=0.042$ and  $\vub=0.0036$.}\label{epsvsVub}
\end{figure}

\begin{figure}[!tb]
 \centering
\includegraphics[width = 0.47\textwidth]{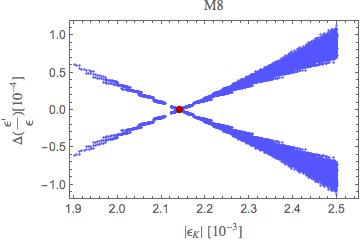}
\includegraphics[width = 0.47\textwidth]{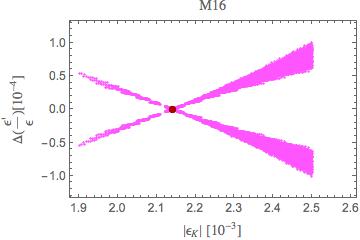}
\caption{ \it $\Delta(\epe)$ versus $\varepsilon_K$ for M8 and M16.
 Red dot represents central SM values. 
$M_{Z^\prime}=10\tev$,  $\vcb=0.042$ and  $\vub=0.0036$.}\label{M810TeV}~\\[-2mm]\hrule
\end{figure}

\begin{figure}[!tb]
 \centering
\includegraphics[width = 0.49\textwidth]{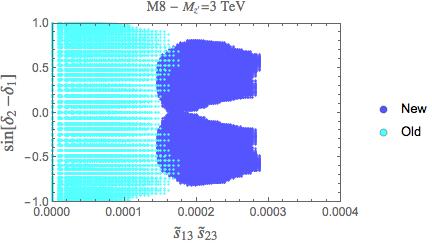}
\includegraphics[width = 0.49\textwidth]{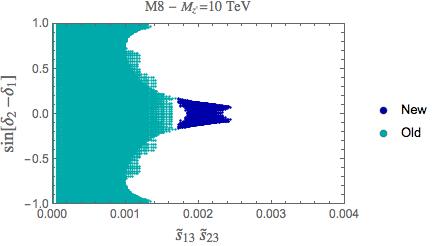}
\caption{ \it Allowed values of $\sin(\delta_2-\delta_1)$ and 
 $\tilde s_{13} \tilde s_{23}$ for M8 and
 $M_{Z^\prime}=3\tev$ (left panel) and  $M_{Z^\prime}=10\tev$ (right panel).
The {\it old} ranges 
are the ones from \cite{Buras:2015kwd} and the {\it new} ones found here.
 $\vcb=0.042$ and  $\vub=0.0036$.
}\label{par310}~\\[-2mm]\hrule
\end{figure}

\boldmath
\subsection{$Z^\prime$ Outside the Reach of the LHC}
\unboldmath
\boldmath
\subsubsection{$\vub=0.042$}
\unboldmath
We will next investigate what happens when higher values of $M_{Z^\prime}$, 
outside the reach of the LHC together with CKM parameters in 
(\ref{CKMfix}), are considered. As an example we set 
 $M_{Z^\prime}=10\tev$. In Fig.~\ref{epsvsVub} we demonstrate that also in this case
the tension between  $\Delta M_{d}$ and  $\varepsilon_K$ can be easily removed.
 In the case of $\Delta M_s$  331 models perform much better than 
the SM represented by the red point so that the inclusion of the uncertainty 
in $F_{B_s}\sqrt{\hat B_{B_s}}$ in (\ref{Kronfeld}), can bring easily 331 
models  
to agree with data which is not possible within the SM.
The 
question then arises what happens with NP effects in other observables for
such high values  $M_{Z^\prime}$. 

On the basis of our discussion in Section~4 in  \cite{Buras:2015kwd} 
we expect the effects in $B_s\to\mu^+\mu^-$ and $C_9$ to be smaller
than for  $M_{Z^\prime}=3\tev$, which can be confirmed as seen in Tables~\ref{panoramaM8} and \ref{panoramaM16}. On the other
hand $\epe$ was found in  \cite{Buras:2015kwd} to be significantly 
enhanced for  $M_{Z^\prime}=10\tev$ as can be seen in Fig.~6 of that paper. 
 Moreover through renormalization group effects it could be even enhanced 
slightly more than for  $M_{Z^\prime}=3\tev$. However, as seen in Fig.~\ref{M810TeV}, with new lattice constraints, this is no longer the case and the 
maximal allowed shifts in $\epe$ are below $1\times 10^{-4}$, far too small to
remove $\epe$ anomaly.

\begin{figure}[!tb]
\centering
\includegraphics[width = 0.47\textwidth]{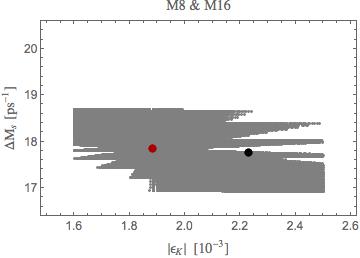}
\includegraphics[width = 0.47\textwidth]{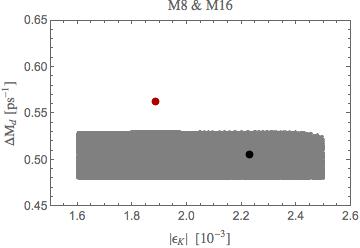}
 \caption{\it  $\Delta M_{s,d}$ vs.  $\varepsilon_K$ in M8 and M16. 
 Red dots represent central SM values and black dots the central experimental values. $M_{Z^\prime}=10\tev$, $\vcb=0.040$ and  $\vub=0.0036$.}\label{MBAJB10TeV4}
\end{figure}

\begin{figure}[!tb]
 \centering
\includegraphics[width = 0.47\textwidth]{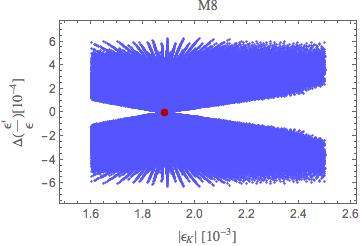}
\includegraphics[width = 0.47\textwidth]{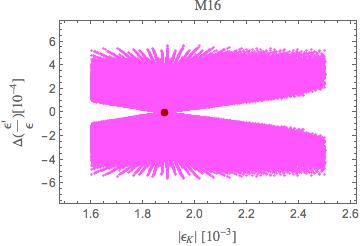}
\caption{ \it $\Delta(\epe)$ versus $\varepsilon_K$ for M8. Red dot represents central SM values.
 $M_{Z^\prime}=10\tev$, $\vcb=0.040$ and  $\vub=0.0036$. }\label{M810TeV4}~\\[-2mm]\hrule
\end{figure}

\begin{figure}[!tb]
 \centering
\includegraphics[width = 0.47\textwidth]{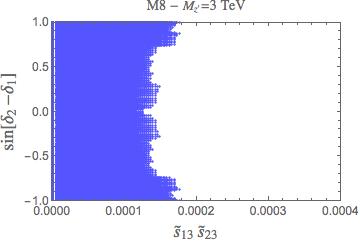}
\includegraphics[width = 0.47\textwidth]{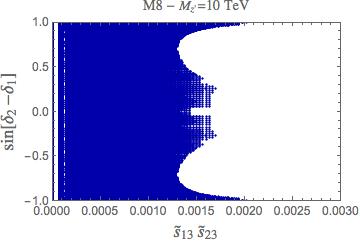}
\caption{ \it Allowed values of $\sin(\delta_2-\delta_1)$ and 
 $\tilde s_{13} \tilde s_{23}$ for M8 and
 $M_{Z^\prime}=3\tev$ (left panel) and  $M_{Z^\prime}=10\tev$ (right panel).
 $\vcb=0.040$ and  $\vub=0.0036$.
}\label{par3104}~\\[-2mm]\hrule
\end{figure}

%%%%%%%%%%%%%%%%%%%%%%%%%%%%%%%%%%%%%%%%%%%%
\begin{table*}[t]
\centering \caption{Summary of results for M8. $\vub=0.0036$.}\label{panoramaM8}
\begin{tabular}{|c   |   c || c| c |c| }\hline
$M_{Z^\prime}$ & $|V_{cb}|$ & $|\Delta(\epe)|_{max} (10^{-4})$ & $\overline{B}(B_{s}\to \mu^+\mu^-)_{min} (10^{-9})$& ${\rm Re}(C_9^{NP})_{min}$  \\
\hline \hline
{$3$ TeV}& $0.042$ & $6.2$ & $3.21$   & $-0.09$
\\  \cline{2-5}
  & $0.040$  &$5.9$ & $2.69$ & $-0.16$
\\ \hline {$10$ TeV}& $0.042$ & $0.98$  & $ 3.45$& $-0.02$
\\  \cline{2-5}
  & $0.040$  & $ 6.94$  & $3.02$ & $-0.05$
  \\ \hline
  \end{tabular}
\end{table*}
%%%%%%%%%%%%%%%%%%%%%%%%%%%%%%%%%%%%%%%%%%%%

%%%%%%%%%%%%%%%%%%%%%%%%%%%%%%%%%%%%%%%%%%%%
\begin{table*}[t]
\centering \caption{Summary of results for M16.  $\vub=0.0036$.}\label{panoramaM16}
\begin{tabular}{|c   |   c || c| c |c| }\hline
$M_{Z^\prime}$ & $|V_{cb}|$ & $|\Delta(\epe)|_{max} (10^{-4})$ & $\overline{B}(B_{s}\to \mu^+\mu^-)_{min} (10^{-9})$& ${\rm Re}(C_9^{NP})_{min}$  \\
\hline \hline
{$3$ TeV}& $0.042$ & $5.55$ & $3.40$   & $-0.28$
\\  \cline{2-5}
  & $0.040$  &$5.17$ & $3.01$ & $-0.49$
\\ \hline {$10$ TeV}& $0.042$ & $0.87$  & $ 3.48$& $-0.05$
\\  \cline{2-5}
  & $0.040$  & $5.95 $  & $3.13$ & $-0.15$
  \\ \hline
  \end{tabular}
\end{table*}
%%%%%%%%%%%%%%%%%%%%%%%%%%%%%%%%%%%%%%%%%%%%

In order to understand this drastic change we recall the general formula for
$\epe$ for arbitrary  $M_{Z^\prime}$   in 331 models 
in the absence of $Z-Z^\prime$ mixing which is irrelevant in M8, M9 and M16
\cite{Buras:2015kwd} 
\be\label{eprimeZPlarge}
\left(\frac{\varepsilon'}{\varepsilon}\right)_{Z^\prime}= \pm r_{\varepsilon^\prime} 1.1 \,[\beta f(\beta)]\,
\tilde s_{13} \tilde s_{23} \sin(\delta_2-\delta_1)
\left[\frac{\bei}{0.76}\right]\left[\frac{3\tev}{M_{Z^\prime}}\right]^2
\ee
with the upper sign for $F_1$ and the lower for $F_2$.  $r_{\varepsilon^\prime}$, 
$\beta$ and $f(\beta)$ are $\ord(1)$ so that only the remaining factors 
are of interest to us. Now as discussed in \cite{Buras:2015kwd} with increasing 
$ M_{Z^\prime}$ larger values of $\tilde s_{13}$ and $\tilde s_{23}$ are allowed
by constraints from  $B^0_{s,d}-\bar B^0_{s,d}$ mixing 
\be
\tilde s_{13}^{\rm max}\propto M_{Z^\prime}, \qquad \tilde s_{23}^{\rm max}\propto M_{Z^\prime}\,, \qquad (\Delta M_{s,d}~~{\rm constraints})\,.
\ee

In this manner the  $M_{Z^\prime}$ suppression in (\ref{eprimeZPlarge}) is 
compensated and the fate of $\epe$ depends on the allowed values of $\sin(\delta_2-\delta_1)$ that follow not only from $\Delta M_{s,d}$ constraints but also 
from 
$S_{\psi K_S}$ and $S_{\psi\phi}$ constraints. Our analysis shows that whereas 
in our previous analysis values of $\sin(\delta_2-\delta_1)=1$ were allowed 
this is no longer the case after new lattice results and maximal values of
$\sin(\delta_2-\delta_1)$  are significantly below unity. While this suppression
appears to be roughly compensated by the increase of the product $\tilde s_{13} \tilde s_{23}$\footnote{These parameters must be larger in order to bring down 
the values of $\Delta M_{s,d}$ to agree with data.} for  $M_{Z^\prime}=3\tev$, 
this is no longer the case for  $M_{Z^\prime}=10\tev$ and $\epe$ is strongly suppressed.
This feature is clearly seen in Fig.~\ref{par310}, where the {\it old} ranges 
are the ones from \cite{Buras:2015kwd} and the {\it new} ones found here.

\boldmath
\subsubsection{$\vcb=0.040$}
\unboldmath
We next consider the CKM input in (\ref{CKMfix1}). The results for 
 $M_{Z^\prime}=10\tev$ are shown in Figs.~\ref{MBAJB10TeV4}--\ref{par3104}.
We observe:
\begin{itemize}
\item
The  tensions between $\Delta M_{s,d}$ and  $\varepsilon_K$ can be much easier 
removed than for  $M_{Z^\prime}=3\tev$ because of the increased NP effects in 
$\varepsilon_K$. Comparing Fig.~\ref{MBAJB10TeV4} with Fig.~\ref{MBAJB10TeV} 
we also observe that the agreement with data is better for $\vcb=0.040$.
\item
The upward shift in $\epe$ up to $(6-7)\times 10^{-4}$ in now possible so that
$\epe$ with  $\vcb=0.040$ can probe much higher mass scales than
it is possible for  $\vcb=0.042$ because of other constraints.
\item
The plots in Fig.~\ref{par3104} when compared with those in  Fig.~\ref{par310} 
explain why the NP effects in $\epe$ for $\vcb=0.040$ have a different structure than for $\vcb=0.042$.  $\sin(\delta_2-\delta_1)$ can for  $\vcb=0.040$  
 reach unity even for $M_{Z^\prime}=10\tev$, while this is not possible for
 $\vcb=0.042$.
\end{itemize}

As seen in Tables~\ref{panoramaM8} and \ref{panoramaM16} NP effects in $B_s\to\mu^+\mu^-$ and $C_9$ are suppressed for 
$M_{Z^\prime}=10\tev$ but not as much as for  $\vcb=0.042$.

All our results for M8 for different values of $\vcb$ and $M_{Z^\prime}$ are 
summarized in Table~\ref{panoramaM8}. Very similar results are obtained for
M9. The corresponding results for M16 are summarized in Table~\ref{panoramaM16}.
These tables show 
again how important is the precise determination of $\vcb$ in tree-level 
decays. 

Finally the red curve in Fig.~\ref{MBAJB10TeV} demonstrates that $\Delta(\epe)$ for $\vcb=0.040$ and $M_{Z^\prime}=10\tev$ can for large values of $\vub$ reach values in the ballpark of
$\simeq 8.8 \times 10^{-4}$. This increase relative to $M_{Z^\prime}=3\tev$ is related to 
renormalization group effects as discussed in detail in  \cite{Buras:2015kwd}.

\section{The Simplest 331 Model: M0}\label{sec:3}
\subsection{Preliminaries}
We will next look at the simplest 331 model recently proposed in \cite{Hue:2015mna} in which $\beta=0$. We will denote it by M0. Even if this model fails to remove most of the anomalies
in question, its simplicity invites us to have a closer look at its flavour 
structure. We will list $Z^\prime$ and $Z$ couplings in this model and present formulae for
 $C_9^\text{NP}$ and $C_{10}^\text{NP}$ as well as $\epe$. 
The expressions for $\Delta F=2$ processes and for $B_s\to\mu^+\mu^-$
 as functions of the couplings listed below  can be found in 
\cite{Buras:2014yna,Buras:2012dp,Buras:2013dea} and we will not repeat them 
here. One only has to set $\beta=0$ in that formulae.
In this manner, in contrast to \cite{Hue:2015mna}, we take $Z-Z^\prime$ mixing in all observables automatically into account.

\boldmath
\subsection{$Z^\prime$ Couplings}
\unboldmath
Setting $\beta=0$ in (17) of \cite{Buras:2013dea} we find for 
quark couplings
{\allowdisplaybreaks
\begin{subequations}
\begin{align}
 \Delta_L^{ij}(Z') &=\frac{g_2}{\sqrt{3}} v_{3i}^*v_{3j}\,,\\
 \Delta_L^{ji}(Z') &=\left[ \Delta_L^{ij}(Z')\right]^\star\,, \\
\Delta_L^{d\bar d}(Z') & = \Delta_L^{u\bar u}(Z')=\Delta_V^{d\bar d}(Z')=\Delta_V^{u\bar u}(Z')=-\frac{g_2}{2\sqrt{3}}
\,,\\
\Delta_A^{d\bar d}(Z') & =    \Delta_A^{u\bar u}(Z')= \frac{g_2}{2\sqrt{3}}\\
\label{RH}
\Delta_R^{d\bar d}(Z')& =  \Delta_R^{u\bar u}(Z')=0\,.
\end{align}
\end{subequations}}%
with $v_{ij}$ given in (\ref{vij}).

The diagonal couplings given here are valid for the first two generations of quarks
neglecting very small additional contributions \cite{Buras:2012dp}.
For the third generation there is an additional term which can  be found in 
(63) of  \cite{Buras:2012dp}. It is irrelevant for our analysis of FCNCs but
plays a role in electroweak precision tests \cite{Buras:2014yna}. For $\beta=0$ the diagonal $b$ quark couplings differ from  $d$ and $s$
couplings only by sign:
\be
\Delta_L^{b\bar b}(Z') =\Delta_V^{b\bar b}(Z')=\frac{g_2}{2\sqrt{3}}, \qquad
\Delta_A^{b\bar b}(Z')=-\frac{g_2}{2\sqrt{3}}\,.
\ee

Setting $\beta=0$ in (18) of \cite{Buras:2013dea} 
 we find for lepton couplings
{\allowdisplaybreaks
\begin{subequations}
\begin{align}
\Delta_L^{\mu\bar\mu}(Z')&=\Delta_L^{\nu\bar\nu}(Z')=\Delta_V^{\mu\bar\mu}(Z')=\frac{g_2}{2\sqrt{3}}\,,\\
\Delta_A^{\mu\bar\mu}(Z') &=  \Delta_A^{\nu\bar\nu}(Z')= -\frac{g_2}{2\sqrt{3}}\,,\\
\Delta_R^{\nu\bar\nu}(Z')&=\Delta_R^{\mu\bar\mu}(Z')= 0
\end{align}
\end{subequations}}%
where we have defined
\begin{align}\label{DeltasVA}
\begin{split}
 &\Delta_V^{\mu\bar\mu}(Z')= \Delta_R^{\mu\bar\mu}(Z')+\Delta_L^{\mu\bar\mu}(Z'),\\
&\Delta_A^{\mu\bar\mu}(Z')= \Delta_R^{\mu\bar\mu}(Z')-\Delta_L^{\mu\bar\mu}(Z').
\end{split}
\end{align}
These definitions also apply to other leptons and quarks. All these couplings 
are evaluated for $\mu=M_{Z^\prime}$ with $g_2=0.633$ for $M_{Z^\prime}=3\tev$.

\boldmath
\subsection{$Z$ Couplings}
\unboldmath
The flavour non-diagonal couplings to quarks are generated from $Z^\prime$ 
couplings through $Z-Z^\prime$ mixing
\be
\Delta^{ij}_L(Z)=\sin\xi \, \Delta^{ij}_L(Z^\prime),
\ee
where using the general formula (10) in \cite{Buras:2014yna}  we find for $\beta=0$ 
\be\label{sxi}
\sin\xi=a\frac{c_W}{\sqrt{3}} \left[\frac{M_Z^2}{M_{Z^\prime}^2}\right]= a\, 4.68\times 10^{-4}\left[\frac{3\tev}{M_{Z^\prime}}\right]^2 \,.
\ee
Here
\be\label{basica}
-1\le a=\frac{1-\tan^2\bar\beta}{1+\tan^2\bar\beta}\le 1, \qquad \tan\bar\beta=\frac{v_\rho}{v_\eta}
\ee
with the scalar triplets $\rho$ and 
$\eta$ responsible for the masses of up-quarks and down-quarks, respectively.
Thus for $\tan\bar\beta=1$ the parameter $a=0$ and the  $Z-Z^\prime$ mixing 
vanish in agreement with  \cite{Hue:2015mna}.
On the other hand in the large $\tan\bar\beta$ limit we find $a=-1$ and 
in the low $\tan\bar\beta$ limit one has $a=1$.

The flavour diagonal $Z$ couplings are the SM ones and collected in  \cite{Buras:2013dea}. We evaluate them with 
$g_2=0.652$ and $\sin^2\theta_W=0.23116$ as valid at $\mu=M_Z$.

\boldmath
\subsection{$C_9^\text{NP}$ and $C_{10}^\text{NP}$}\label{RARE}
\unboldmath
The corrections from NP to the Wilson coefficients $C_9$ and $C_{10}$ that weight the semileptonic operators in the effective hamiltonian relevant
for $b\to s\mu^+\mu^-$ transitions 
are given as follows
\begin{align}
 \sin^2\theta_W C^{\rm NP}_9 &=-\frac{1}{g_{\text{SM}}^2M_{Z^\prime}^2}
\frac{\Delta_L^{sb}(Z')\Delta_V^{\mu\bar\mu}(Z')} {V_{ts}^* V_{tb}}(1+R^V_{\mu\mu}) ,\label{C9}\\
   \sin^2\theta_W C^{\rm NP}_{10} &= -\frac{1}{g_{\text{SM}}^2M_{Z^\prime}^2}
\frac{\Delta_L^{sb}(Z')\Delta_A^{\mu\bar\mu}(Z')}{V_{ts}^* V_{tb}}(1+R^A_{\mu\mu})\label{C10}.
 \end{align}
As seen in these equations $C^{\rm NP}_9$ involves leptonic vector coupling 
of $Z^\prime$ while $C^{\rm NP}_{10}$ the axial-vector one. $C^{\rm NP}_9$ is 
crucial for $B_d\to K^*\mu^+\mu^-$, $C^{\rm NP}_{10}$ for $B_s\to\mu^+\mu⁻$ and 
both coefficients are relevant for $B_d\to K\mu^+\mu^-$.

Here
\be\label{gsm}
g_{\text{SM}}^2=
4 \frac{M_W^2 G_F^2}{2 \pi^2} = 1.78137\times 10^{-7} \gev^{-2}\,,
\ee
with $G_{F}$ being the Fermi constant. The terms $R^V_{\mu\mu}$ and 
$R^A_{\mu\mu}$ are generated by $Z-Z^\prime$ mixing and are given as follows

\be\label{DF1}
R^V_{\mu\mu}=\sin\xi \left[\frac{M_{Z\prime}^2}{M_{Z}^2}\right] \left[\frac{\Delta^{\mu\mu}_V(Z)}{\Delta^{\mu\mu}_V(Z^\prime)}\right]\,,
\ee
\be\label{DF2}
R^A_{\mu\mu}=\sin\xi \left[\frac{M_{Z\prime}^2}{M_{Z}^2}\right] \left[\frac{\Delta^{\mu\mu}_A(Z)}{\Delta^{\mu\mu}_A(Z^\prime)}\right]\,,
\ee

For $\beta=0$ we find then 
\begin{align}
 \sin^2\theta_W C^{\rm NP}_9 &=-\frac{1}{g_{\text{SM}}^2M_{Z^\prime}^2}
\frac{g_2^2(M_{Z^\prime})}{6}
\left[\frac{v_{32}^*v_{33}} {V_{ts}^* V_{tb}}\right](1+R^V_{\mu\mu}) ,\label{C90}\\
   \sin^2\theta_W C^{\rm NP}_{10} &= +\frac{1}{g_{\text{SM}}^2M_{Z^\prime}^2}\frac{g_2^2(M_{Z^\prime})}{6} \left[\frac{v_{32}^*v_{33}} {V_{ts}^* V_{tb}}\right](1+R^A_{\mu\mu})\label{C100}
 \end{align}
with
\be
R^V_{\mu\mu}=-0.08\, a, \qquad 
R^A_{\mu\mu}=-1.02 \, a\,
\ee
that do not depend on $M_{Z^\prime}$ except for logarithmic $M_{Z^\prime}$ dependence of $g_2$. The numerical factors above correspond to $M_{Z^\prime}=3\tev$.

We observe then that in the absence of $Z-Z^\prime$ mixing ($a=0$), independently of the parameters $v_{ij}$, the following phenomenologically  successful 
relation
\be\label{R1a}
C_9^\text{NP}=-C_{10}^\text{NP},  \qquad (a=0)
\ee
holds.  This should be contrasted with models M8, M9 and M16 
for which we found \cite{Buras:2015kwd}
\be\label{M8M9}
 C_9^{\rm NP}=0.49\, C_{10}^{\rm NP}\quad ({\rm M8})\,, \qquad 
 C_9^{\rm NP}=0.42\, C_{10}^{\rm NP}\quad ({\rm M9})\,.
\ee

The result in (\ref{R1a}) differs also from 
\be\label{M16}
C_9^{\rm NP}=-4.59\, C_{10}^{\rm NP}\quad ({\rm M16}) 
\ee
which is close to one of the favourite solutions in which NP resides dominantly in the coefficient $C_9$. 
Thus already on the basis of $B$ physics observables we should be able to 
distinguish between the models M0, (M8,M9) and M16.

However in the presence of  $Z-Z^\prime$ mixing the relation (\ref{R1a}) does not
hold. While this effect is small in $C_9^{\rm NP}$, it can be large in $C_{10}^{\rm NP}$, in particular for $a=1$, when $C_{10}^{\rm NP}$ becomes very small and the suppression 
of the rate for $B_s\to\mu^+\mu^-$ is absent. More interesting is then the 
case of $a\approx -1$, corresponding to large $\tan\bar\beta$, as then the simultaneous suppressions of $C_9$ through $C_9^{\rm NP}$
 and of $B_s\to\mu^+\mu^-$ rate through $C_{10}^{\rm NP}$ are stronger. We find
then
\be\label{R2a}
C_9^\text{NP}\approx -0.5\, C_{10}^\text{NP}, \qquad (a\approx -1),
\ee
that on a qualitative level is still  a  better description of the data than the results in (\ref{M8M9}) and (\ref{M16}). But the crucial question is whether 
the values of both coefficients are sufficiently large when all constraints are
taken into account. Before answering this question let us make a closer look 
at $\epe$ in this model.

\boldmath
\subsection{$\epe$}
\unboldmath
\subsubsection{Preliminaries}
The analyses of $\epe$ in 331 models with $\beta\not=0$ have been 
presented by us in \cite{Buras:2014yna,Buras:2015kwd}. We want to generalize 
them to the case $\beta=0$. Generally in 331 models we have
\be\label{total}
\left(\frac{\varepsilon'}{\varepsilon}\right)_{331}=\left(\frac{\varepsilon'}{\varepsilon}\right)_{\rm SM}+\left(\frac{\varepsilon'}{\varepsilon}\right)_{Z}+
\left(\frac{\varepsilon'}{\varepsilon}\right)_{Z^\prime}\equiv \left(\frac{\varepsilon'}{\varepsilon}\right)_{\rm SM}+\Delta(\epe)
\, 
\ee
with the $\Delta(\epe)$  resulting from tree-level $Z^\prime$ and $Z$ exchanges.

Now, as demonstrated by us in  \cite{Buras:2014yna}, the shift  $\Delta(\epe)$ 
is governed in 331 models 
by the electroweak $(V-A)\times (V+A)$ penguin operator 
\begin{equation}\label{O4} 
Q_8 = \frac{3}{2}\,(\bar s_{\alpha} d_{\beta})_{V-A}\!\!\sum_{q=u,d,s,c,b,t}
      e_q\,(\bar q_{\beta} q_{\alpha})_{V+A} \,
\end{equation}
with only small contributions from other operators. This result applies to both 
$Z$ and $Z^\prime$ contributions with the latter ones significantly more important as demonstrated in  \cite{Buras:2014yna}. Here we would like to point out 
that this pattern is not valid for $\beta=0$.  

Indeed as seen in (41) of  \cite{Buras:2014yna} the important coefficient 
$C_7(M_{Z^\prime})$ generated by tree-level $Z^\prime$ exchange for $\beta\not=0$ vanishes for $\beta=0$ and consequently,  $Q_8$ operator cannot 
be generated from $Q_7$ operator by renormalization group effects. Contributions of other operators are very small  
 so that $Z^\prime$ contributions to $\epe$ can be neglected. This is directly 
related to the fact, as seen in (\ref{RH}), that the diagonal right-handed couplings of $Z^\prime$ to
quarks vanish for $\beta=0$. But for $Z$ such couplings are present implying 
that tree-level $Z$ exchanges can provide a shift in $\epe$.

\boldmath
\subsubsection{$Z$ Contribution}
\unboldmath
The inclusion of this contribution  is straightforward  as the only thing to be done is to calculate the shifts from NP
in the functions $X$, $Y$ and $Z$ that enter the SM model contribution to $\epe$. One finds then \cite{Buras:2014yna}
\be\label{DXYZ}
\Delta X=\Delta Y =\Delta Z= \sin\xi \, c_W\frac{8\pi^2}{g_2^3}\frac{{\rm Im}\Delta_L^{sd}(Z^\prime)}{{\rm Im}\lambda_t}
\ee
where $g_2=0.652$ and $\lambda_t=V_{td}V^*_{ts}$. Replacing then the SM functions 
$X_0(x_t)$, $Y_0(x_t)$ and $Z_0(x_t)$ by 
\be
X=X_0(x_t)+\Delta X, \qquad Y=Y_0(x_t)+\Delta Y, \qquad Z=Z_0(x_t)+\Delta Z
\ee
in the phenomenological formula formula (90) for $\epe$ in 
 \cite{Buras:2015yba}  allows to take automatically the first two 
contributions in (\ref{total})  in 331 models into account. 

Inserting $\sin\xi$ in (\ref{sxi}) and the $Z^\prime$ coupling into (\ref{DXYZ}) we obtain for $\beta=0$ 
\be\label{DXYZ0}
\Delta X=\Delta Y =\Delta Z= 47.6\, a\, \left[\frac{M_Z^2}{M_{Z^\prime}^2}\right] \frac{{\rm Im}(v_{32}^*v_{31})}{{\rm Im}\lambda_t}.
\ee
Evidently for $a=0$, as done in  \cite{Hue:2015mna}, NP contributions to 
$\epe$ from $Z$ exchanges vanish and as the ones from $Z^\prime$ can be
neglected, $\epe$ is full governed by the SM contribution. This appears 
presently problematic 
in view of the findings in  \cite{Bai:2015nea,Buras:2015yba,Buras:2015xba, Buras:2015jaq} that a significant upwards shift $\epe$ of at least $5\times 10^{-4}$ 
is required to bring the theory to agree with the data from 
from NA48 \cite{Batley:2002gn} and KTeV
\cite{AlaviHarati:2002ye,Abouzaid:2010ny} collaborations.

The question then arises whether including  $Z-Z^\prime$ mixing 
we can obtain the required positive shift in $\epe$. But,  as seen in 
(\ref{R2a}), in order to preserve 
at least partly the pattern in (\ref{R1a})  we are interested in 
\be\label{arange}
-1\le a < 0, \qquad \tan\bar\beta > 1\,.
\ee

In order to answer this question 
we insert 
(\ref{DXYZ0}) into (90) in Appendix B of \cite{Buras:2015yba} 
to obtain first
\be
\Delta(\epe)=\Delta(\epe)_Z= 47.6 \, a\left[\frac{M_Z^2}{M_{Z^\prime}^2}\right][P_X+P_Y+P_Z] {\rm Im}(v_{32}^*v_{31}).
\ee
From Table 5 in \cite{Buras:2015yba} we find then for the central value of 
$\alpha_s(M_Z)$:
\be
P_X+P_Y+P_Z= 1.52+0.12\, R_6-13.65\, R_8
\ee
where
\be\label{RS}
R_6\equiv \bsi\left[ \frac{114.54\mev}{m_s(m_c)+m_d(m_c)} \right]^2,\qquad
R_8\equiv \bei\left[ \frac{114.54\mev}{m_s(m_c)+m_d(m_c)} \right]^2.
\ee
Now the results from  the RBC-UKQCD collaboration  imply the following values 
for $\bsi$ and $\bei$ \cite{Buras:2015yba,Buras:2015qea}
\be\label{Lbsi}
\bsi=0.57\pm 0.19\,, \qquad \bei= 0.76\pm 0.05\,, \qquad (\mbox{RBC-UKQCD})
\ee
 that are compatible with  the bounds from large $N$ approach 
 \cite{Buras:2015xba}
\be\label{NBOUND}
\bsi\le \bei < 1 \, \qquad (\mbox{\rm large-}N).
\ee
Using then the results in (\ref{Lbsi})
we find 
\be
P_X+P_Y+P_Z= -8.78\pm 0.68
\ee
and finally 
\be
\Delta(\epe)=-a\, (0.39\pm 0.03)\left[\frac{3\tev}{M_{Z^\prime}}\right]^2 {\rm Im}(v_{32}^*v_{31}).
\ee
The question then arises whether for $a$ in the range (\ref{arange}) one 
can get sufficient shift in $\epe$ while satisfying other constraints. 
In particular the ones from $\Delta F=2$ transitions, where as seen in the 
previous section the SM experiences tensions in 
its predictions for $\Delta M_{s,d}$ and $\varepsilon_K$. 

\begin{figure}[!tb]
\centering
\includegraphics[width = 0.47\textwidth]{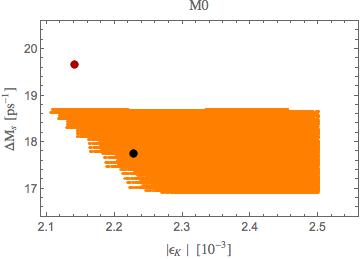}
\includegraphics[width = 0.47\textwidth]{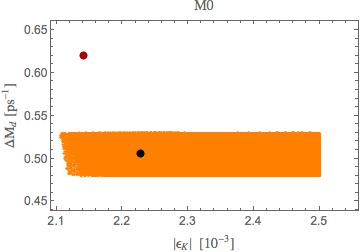}
 \caption{\it  $\Delta M_{s,d}$ vs.  $\varepsilon_K$ in M0. 
 Red dots represent central SM values and black dots the central experimental values.  $M_{Z^\prime}=3\tev$ and $\vcb=0.042$.}\label{MBAJB0}
\end{figure}

\begin{figure}[!tb]
 \centering
\includegraphics[width = 0.60\textwidth]{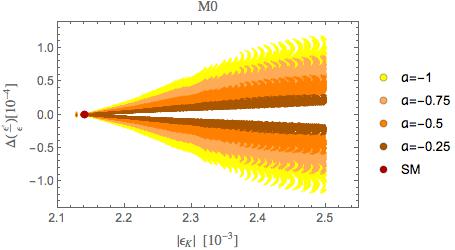}
\caption{ \it $\Delta(\epe)$ versus $\varepsilon_K$ for M0
 for several values of the $Z-Z^\prime$ mixing parameter $a$.  $M_{Z^\prime}=3\tev$ and $\vcb=0.042$.}\label{M0}~\\[-2mm]\hrule
\end{figure}

In Fig.~\ref{MBAJB0} we show $\Delta M_{s,d}$ vs. $\varepsilon_K$ in M0. Red dots represent central SM values and black dots the central experimental 
values.  We observe that 
the tensions between   $\Delta M_{s,d}$ vs.  $\varepsilon_K$ present in the 
SM  can be removed in the M0 model. But as seen in Fig.~\ref{M0} the shift in 
$\epe$ can be at most $1.1\times 10^{-4}$ which is far too small to be able
to remove $\epe$ anomaly. Moreover, this maximal shift can only be obtained 
for the maximal $Z-Z^\prime$ mixing. We have checked using the expressions 
in \cite{Buras:2014yna} that then  the fit to EWPO is significantly 
worse than the one in the SM, whereas the three models analysed in the previous
section perform better in these tests than the SM  \cite{Buras:2014yna}.

We do not show the results for $C_9$ and $B_s\to\mu^+\mu^-$ as NP effects 
are significantly smaller than in M8 and M16. Thus even if  M0 can remove 
the  tensions between $\Delta M_{s,d}$ and $\varepsilon_K$, it fails badly 
in the case of other anomalies and therefore cannot compete with models 
M8, M9 and M16, unless all anomalies disappear one day.

\section{Summary}\label{sec:4}
Motivated by the recently improved results from the Fermilab Lattice and MILC Collaborations  on the hadronic matrix elements entering  $\Delta M_{s,d}$ in
 $B_{s,d}^0-\bar B_{s,d}^0$ mixing \cite{Bazavov:2016nty} 
and the resulting increased tensions 
between $\Delta M_{s,d}$  and $\varepsilon_K$ in the SM and 
generally CMFV models \cite{Blanke:2016bhf}, we have performed a new analysis of 331 models.
In order to illustrate the sensitivity of the results to the modification of 
hadronic parameters we have first used the CKM input of our previous analysis 
in  \cite{Buras:2015kwd} that is given in  (\ref{CKMfix}). In addition, in order
to illustrate the sensitivity of the results to the value  of $\vcb$ 
we have also performed the analysis with the CKM input in (\ref{CKMfix1}), 
where $\vcb$ is lower than in (\ref{CKMfix}).  We also investigated $\vub$ 
dependence.

The most important results of our analysis, summarized in Tables~\ref{panoramaM8} and \ref{panoramaM16} are as follows:
\begin{itemize}
\item
The tensions between $\Delta M_{s,d}$  and $\varepsilon_K$ 
can be removed in  the three 331 models with $\beta\not=0$ (M8, M9, M16) considered by us and this for both CKM inputs. This turns out to be also possible in the model with $\beta=0$ (M0) in the case of the input  (\ref{CKMfix}) but it is 
much harder when $\vcb$ is smaller as in  (\ref{CKMfix1}).
\item
Models M8, M9 and M16  can provide a positive shift in $\epe$ up to $6\times 10^{-4}$ for
 $M_{Z^\prime}=3\tev$ for both choices of $\vcb$  and $\vub=0.0036$. But in contrast to our previous analysis this shift decreases fast with increasing  $M_{Z^\prime}$ in the case
of $\vcb=0.042$ but its maximal values are practically unchanged for 
 $M_{Z^\prime}=10\tev$ when $\vcb=0.040$  is used. We also find that for  $\vcb=0.040$ and the 
inclusive values of $\vub$ the maximal shifts in $\epe$ are increased to 
 $7.7 \times 10^{-4}$ and $8.8\times 10^{-4}$ for  $M_{Z^\prime}=3\tev$ and 
 $M_{Z^\prime}=10\tev$, respectively.
In the 
model M0, in which NP contribution to $\epe$ is governed by $Z-Z^\prime$ mixing, NP effects are very small even for  $M_{Z^\prime}=3\tev$.
\item
In M8 and M9 the rate for $B_s\to\mu^+\mu^-$ can be reduced by at most $10\%$ 
and $20\%$ for  $M_{Z^\prime}=3\tev$  and  $\vcb=0.042$ and  $\vcb=0.040$, 
respectively. This can bring the theory within $1~\sigma$ range of
the combined result from CMS and LHCb and for  $\vcb=0.040$ one can even 
reach the present central experimental value of this rate (\ref{LHCb2}). The maximal 
shifts in $C_9$ are 
$C_9^\text{NP}=-0.1$ and $C_9^\text{NP}=-0.2$ for these two $\vcb$ values, 
respectively. This is only a moderate shift and these models 
do not really help 
in the case of $B_d\to K^*\mu^+\mu^-$ anomalies.
\item
In M16 the situation is opposite. The rate for $B_s\to\mu^+\mu^-$ can be reduced for  $M_{Z^\prime}=3\tev$ for the two $\vcb$ values by at most $3\%$ and $10\%$, 
respectively but 
  with the corresponding values $C_9^\text{NP}=-0.3$ and $-0.5$ the anomaly  in $B_d\to K^*\mu^+\mu^-$ can be partially reduced.
\item
In M0 NP effects in $\epe$, $B_s\to\mu^+\mu^-$ and $B_d\to K^*\mu^+\mu^-$
are too small to be relevant.  Therefore our analysis demonstrates that
in the presence of the anomalies discussed
by us the $U(1)_X$ factor in the gauge group
of 331 models cannot be  $U(1)_Y$ .
\item
For higher values of  $M_{Z^\prime}$ the effects in $B_s\to\mu^+\mu^-$ and $B_d\to K^*\mu^+\mu^-$ are much smaller.
We recall that  NP effects in rare $K$ decays and $B\to K(K^*)\nu\bar\nu$ remain small in all 331 models even for  $M_{Z^\prime}$ of few TeV.
\end{itemize}

 Even if  models M8, M9, M16 still compete with each other and M0 
does not appear to be phenomenologically viable from present perspective, 
our feeling is that eventually only models M8 and M9 have a chance to survive future tests if the anomalies discussed by us will be confirmed in the future. The point is that with present theoretical uncertainties in 
 $B_d\to K^*\mu^+\mu^-$ NP effects, even in M16, will be hardly seen in this 
decay. The decay $B_s\to\mu^+\mu^-$ is much cleaner and in the flavour
precision era $15-20\%$ effects from NP, which are only possible in M8 and M9, 
could in principle be distinguished from SM predictions but this would require
very large reduction in the experimental error on its rate. 

Thus the main virtue 
of 331 models as opposed to SM and CMFV models is the ability to remove
the tensions between $\Delta M_{s,d}$  and $\varepsilon_K$ and simultaneously 
provide a significant upward shift in $\epe$ but only for lower values of 
$\vcb$ can this property remain for  $M_{Z^\prime}$ beyond the LHC reach. 
The possibility of a significant suppression of the rate for $B_s\to\mu^+\mu^-$ in M8 and M9 for $\vcb=0.040$ is also a welcome feature.  In particular, as 
it is correlated with the maximal shift in $\epe$.

 While the NP pattern  in flavour physics identified by us in 331 models 
is interesting, we should hope that eventually  NP 
contributions to flavour observables are larger than found in these models and 
are also significant  in rare $K$ decays which are theoretically very clean and in  $B\to K(K^*)\nu\bar\nu$ which are cleaner than $B\to K(K^*)\mu^+\mu^-$ decays.
 Most importantly the comparison of our results in \cite{Buras:2015kwd}, 
prior to the lattice results in  \cite{Bazavov:2016nty},
 with 
the ones obtained using this new input  demonstrates clearly how the shifts and increased accuracy
in non-perturbative parameters can have important impact on the size of NP
effects. Similar comment can be made in connection with $\vcb$.

\section*{Acknowledgements}
This research was done and financed in the context of the ERC Advanced Grant project ``FLAVOUR''(267104) and has also been carried out within the INFN project (Iniziativa Specifica) QFT-HEP. It  was partially
supported by the DFG cluster
of excellence ``Origin and Structure of the Universe''.

\bibliographystyle{JHEP}
\bibliography{allrefs}
\end{document}